\documentclass[pre,twocolumn,showpacs,superscriptaddress]{revtex4}
\usepackage{epsfig}

\begin{document}   

\title{Langevin theory of absorbing phase transitions with a conserved magnitude}

\author{Jos\'e J. Ramasco}\email{jjramasc@fc.up.pt}\affiliation{Departamento 
de F\'{\i}sica and Centro de F\'{\i}sica do Porto,\\
Faculdade de Ci\^{e}ncias, Universidade do Porto,\\
Rua do Campo Alegre 687, 4169-007 Porto, Portugal.}  

\author{Miguel A. Mu\~noz}\email{mamunoz@onsager.ugr.es}
\affiliation{Instituto de F\'{\i}sica Te\'orica y Computacional Carlos I,\\
Universidad de Granada, Facultad de Ciencias, 18071 Granada, Spain.}

\author{Constantino A. da Silva Santos}\affiliation{Departamento 
de F\'{\i}sica and Centro de F\'{\i}sica do Porto,\\
Faculdade de Ci\^{e}ncias, Universidade do Porto,\\
Rua do Campo Alegre 687, 4169-007 Porto, Portugal.}

\date{\today}                  
    
\begin{abstract}  
The recently proposed Langevin equation, aimed to capture the relevant critical 
features of stochastic sandpiles, and other self-organizing systems is 
studied numerically. 
This equation is similar to the Reggeon field theory, describing
generic systems with absorbing states, but it is coupled linearly to a second
conserved and static (non-diffusive) field. It has been claimed to represent
a new universality class, including different discrete models: 
the Manna as well as other sandpiles, reaction-diffusion systems, etc.
In order to integrate the equation, and surpass the difficulties associated with
its singular noise, we follow a numerical technique introduced by Dickman. 
Our results coincide remarkably well with those of discrete models 
claimed to belong to this universality class, in one, two, and
three dimensions. 
This provides a strong backing for the Langevin theory of stochastic sandpiles, 
and to the very existence of this new, yet meagerly understood, universality class.
\end{abstract}   


\maketitle 

 Aimed at shedding some light at the origin of ``{\it order}'' in Nature, 
some different routes to organization have been proposed in the last 
fifteen years or so.
In particular, the concept of {\it self-organization}, as exemplified by   
{\it sandpiles} \cite{btw,manna,zhang1} (for reviews see \cite{Jensen,GG,BJP})
(one of the canonical instances of self-organizing systems) has generated
a rather remarkable outburst of interest. 
In order to rationalize sandpiles in particular, and {\it self-organized criticality} 
(SOC) in general, and to understand their critical properties, it has been 
recently proposed to look at them as systems with many absorbing states
\cite{BJP,vdmz,dvz,fes,1d}.
The underlying idea is that in the absence of external driving sandpile models get 
eventually trapped into stable configurations
from which they cannot escape, {\it i.e.} absorbing states (AS) \cite{MD,Granada,Hinrichsen}.
In order to make the aforementioned connection more transparent, 
the notion of {\it fixed energy sandpiles} (FES) was introduced.
These modified sandpiles are defined in such a way that they share the microscopic 
rules with their standard (slowly driven) counterparts, but with no driving (no 
addition of sand-grains) nor dissipation; {\it i.e.} the total 
amount of sand (energy) becomes a conserved quantity that acts as a free 
tunable parameter. 
In this way, if a standard sandpile in its stationary, critical state has a 
density of grains (or energy) $\zeta_c$, it can
be verified that its fixed-energy counterpart exhibits a transition from
an active to an absorbing phase at precisely $\zeta_c$, while
it is in an absorbing (active) state below (above) that energy density.
Slow driving and dissipation define a mechanism able to pin the system to its 
critical point \cite{BJP,vdmz,dvz,fes,1d}. 

Using this analogy with systems with AS, a field theoretical description of {\it stochastic 
sandpiles} has been proposed\cite{BJP,vdmz,dvz,fes}, which 
includes the two more relevant features of stochastic sandpiles: 
{\bf i}) {\it the presence of infinitely many AS}
 and, {\bf ii}) {\it the global conservation of the total energy}.
The phenomenological field theory (Langevin equation) aimed to capture the 
relevant critical features of this type of systems is similar to the well known
Reggeon field theory (RFT) \cite{conjecture,Granada,Hinrichsen} (describing generic
systems with AS) but it is 
coupled linearly to a conserved energy field, namely \cite{vdmz,fes}:
\begin{equation}
\label{fesp}  
\begin{array}{ll}
\partial_t \rho = & D_a \nabla^2 \rho - \mu \rho - \lambda \rho^2 + \omega \rho \phi +
\eta(x,t) \\
\partial_t \phi = & D_c \nabla^2 \rho \mbox{ ,}
\end{array}
\end{equation}
\noindent
where $D_a$, $D_c$, $\mu$, $\lambda$ and $\omega$ are constants, $\rho(x,t)$ and
 $\phi(x,t)$ are the activity and the energy field respectively, 
$\eta$ is a zero-mean Gaussian noise with:
\begin{equation}
\label{corr}
\langle \, \eta(x,t)\, \eta(x',t') \, \rangle = 2\, \sigma^2 \, \rho(x,t)\, 
\delta(x-x')\, \delta(t-t') .
\end{equation}
\noindent
As usual in systems with AS, the variance of $\eta$ is proportional 
to $\rho$  \cite{conjecture,Hinrichsen}.

 Soon after the introduction of the previous Langevin equation 
its range of applicability was extended, as it was conjectured 
to describe all systems with many AS 
and an auxiliary conserved and non-diffusive (or static) field \cite{Rossi}.
In particular for a reaction-diffusion model in this family an 
equation similar to Eq.(\ref{fesp}) was 
derived rigorously by using standard Fock-space formalism techniques
\cite{Rossi,Romu1,Romu2}. To be more precise, we should mention that the
derived set of equations, includes some higher order terms, as noise
crossed-correlations, whose role in the asymptotic
properties is unclear. 

A priori, it is not straightforward to decide from a field theoretical 
point of view whether the extra conservation law induces
a critical behavior different from that of RFT or if, on the contrary, 
it is an irrelevant perturbation at the RFT renormalization group 
fixed point \cite{conjecture}.
From the theory  side, it has been recently argued by van Wijland 
that the Langevin equation is renormalizable in $d_c=6$ \cite{fred}, while 
other authors have previously claimed $d_c=4$ \cite{vdmz,fes,BJP}. 
Some mean field results and simulations in high dimensions of discrete
models \cite{Lubeck}, and also a new method
recently proposed by Lubeck to determine the upper critical dimension 
of systems with AS \cite{Lubeck2} leads rather convincingly to $d_c=4$, but
we  are still far from a full clarification of these issues at a 
field theoretical level. In any case, it is commonly accepted, from
numerical evidence, that this constitutes a new universality  class, usually
called  Manna-class, or C-DP (in the spirit of Hohenberg and Halperin \cite{HH})
\cite{fes,BJP,Rossi,Romu1,Romu2,chate,Lubeck,Lubeck2}. 

In order to clarify the situation, it is the purpose of this paper to integrate 
numerically Eq.(\ref{fesp}) in $1$, $2$, and $3$ dimensions. 
In this way we will verify whether this set of Langevin equations 
(without the higher order terms  obtained in \cite{Romu1}) 
describes correctly the critical properties of the discrete models reported 
to belong to this class. 

With this purpose we employ an integration scheme introduced by
Dickman some years back \cite{Method} which, to the best  of our knowledge, is 
the only working method for Langevin equations including a RFT-like type
of noise ({\it i. e. } with variance linear in the activity field).
We will verify that indeed the Langevin equation as it is reproduces remarkably 
well the known exponents (as measured in discrete models in this class), thus
providing us with a sound base for further theoretical analyses of this 
universality class and of the role of conservation in self-organizing
systems \cite{GG}.

{\it The model.--}  
We integrate numerically Eq.(\ref{fesp}) \cite{vdmz,fes,BJP}. 
A technical problem appears when a standard (Euler \cite{Euler}) discretization 
scheme is used: due the symmetry of $\eta$ around zero, and the 
fact that the noise-term dominates the evolution
whenever the density field is sufficiently small, 
negative (unphysical) local values of the density field can be generated.

In order to overcome this difficulty a different, non-trivial, 
integration scheme was proposed by Dickman for the RFT \cite{Method}.    
It consists in discretizing the density field $\rho$ as well as time and
space. The {\it quanta} of density of activity can be taken 
proportional to the discrete time step, $\Delta \rho = \Delta t$, 
in such a way that the continuity of $\rho$ is recovered in 
the thermodynamic limit ($\Delta t, \Delta x \rightarrow 0$).
The activity density at a given site $i$ and time $t$ is then 
given by $\rho(i,t) = m(i,t) \Delta\rho$, where $m(i,t)$ takes integer values. 
Further details of the scheme can be found 
in \cite{Method}. It has been successfully applied to both the RFT \cite{Method} 
and to systems with many AS \cite{Clopez}, leading to 
good estimations of phase diagrams and critical properties.
In order to extend the algorithm to our problem, the second equation of
(\ref{fesp}) is integrated using an usual Euler scheme with a continuous 
field $\phi(x,t)$. While for the equation of $\rho$, we follow a subtle 
different procedure based on Dickman's method. First, we calculate  
\begin{eqnarray}
& f(i,t+\Delta t)-f(i,t) =  \Delta t  [D_a  \nabla_d^2  m(i,t) - 
\mu  m(i,t) \nonumber \\ 
& - \lambda  \Delta\rho\, m^2(i,t) + \omega  m(i,t)  \phi(i,t)] 
 +  \sigma  m^{1/2}(i,t)  \eta'(i,t)
\end{eqnarray} 
\noindent
where $\Delta t = \Delta\rho$, $\eta'$ is a zero-mean 
Gaussian white noise, $\nabla^2_d$ is the discrete Laplacian operator and 
$f(x,t)$ is an auxiliary continuous field. 
Then, after each integration step, the number of quanta of $\rho$, $m(x,t)$, is
updated according to:
\begin{equation}\begin{array}{l}
m(x,t+\Delta t) \rightarrow m(x,t) + \mbox{int}[f(x,t+\Delta t)] \\
\,\\
f(x,t+\Delta t) \rightarrow f(x,t+\Delta t) - \mbox{int}[f(x,t+\Delta t)] .
\end{array}
\end{equation}
Initial conditions are taken as follows: {\bf i)}~~
$\phi(x,t=0) = \phi_0 \times \, (1 + a \nabla_d^2 \, \varepsilon(x,t))$,
where $\varepsilon$ is a normalized Gaussian noise with zero average, 
and $a$ is a constant  establishing the range of
relative variation allowed to $\phi$ with respect to its mean value, $\phi_0$.
$\phi_0$, is the control parameter, and except for transient effects results should 
not depend on $a$. {\bf ii)} The initial condition for
$\rho$ is chosen by randomly 
distributing  active-field quanta, in such a way that $\rho(x,t=0) \le \phi(x,t=0)$
everywhere. 

We have carried out extensive simulations of the coupled Eqs.(\ref{fesp}) in one,
two and three dimensional lattices. In all the cases, the time-mesh has
been fixed to $\Delta t = 0.01$, and $\Delta x=1$,
(values below which we have verified that results are not further affected). 
We also fix $D_a = D_c = 5$ and $\mu = \lambda = \omega = a = 1$. The noise 
amplitude $\sigma$ is taken different for the various dimensions in order to
fix the transition in a reasonable (but arbitrary) value 
of $\phi_0$: $\sigma = 1$ in $1D$, $\sigma = 0.5$ in $2D$, and $\sigma = 0.35$ in $3D$.   
We have verified that the total energy is conserved within the considered 
precision, in all cases. The number of runs goes from $10^2$ up to $10^5$ 
depending on system size. 

{\it Results.--} As we vary $\phi_0$, a continuous transition separating the
absorbing from active phase, is observed at a critical threshold $\phi_c$.
Below (above) this value the system is absorbing (active) states. 
The usual scaling laws:  $\rho \sim (\phi_0-\phi_c )^\beta$, 
$\xi\sim (\phi_0-\phi_c )^{-\nu_\perp}$, 
$\tau \sim (\phi_0-\phi_c)^{-\nu_\parallel}$, where $\xi$ ($\tau$) is the 
correlation length (time) are expected to hold \cite{MD,Hinrichsen,Granada}.
 This leads to the definition of the dynamic exponent as $\tau \sim \xi^z$, with 
$z = \nu_\parallel/\nu_\perp$. It is also expected that at the critical point
the density of activity presents a power law decay with time, 
$\rho \sim t^{-\theta}$. However, in some models, an anomalous 
critical time behavior of $\rho$ has been reported \cite{fes,1d,Rossi,Romu1,Romu2}. 
We shall later return to this issue.
\begin{figure}
\begin{center}
\epsfxsize=6.5cm
\epsfbox{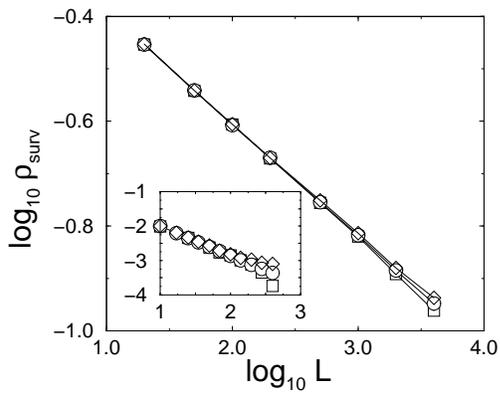}
\caption{Stationary value of $\rho_{surv}$ for different system sizes in $1D$.
Squares correspond to $\phi_0 = 1.6369$, 
circles to $\phi_0 = 1.6371$ and diamonds to $\phi_0 = 1.6373$. In the inset, the
same graph is displayed but for $2D$ data; 
squares are for $\phi_0 = 0.631$, circles for $\phi_0 =
0.6325$ and diamonds for $\phi_0 = 0.635$.}
\end{center}
\end{figure} 
As usual the finite size of simulated systems induces the possibility of falling 
into the AS even for $\phi_0 > \phi_c$. 
This fact has two consequences. The density of activity $\rho$ does not reach a
stationary state close to the critical point. Hence, we are forced 
to consider the density of surviving trials, $\rho_{surv}$, in order to 
realize the finite size analysis of $\rho$. 
 On the other hand, this provides us with a method to measure the dynamic 
exponent $z$, by determining a characteristic decaying time as a function
of system size.
\begin{figure}
\begin{center}
\epsfxsize=6.5cm
\epsfbox{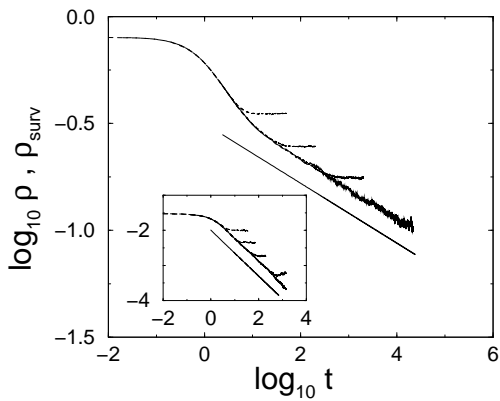}
\caption{Evolution of $\rho$ (continuous line) and $\rho_{surv}$
(dashed lines) for several system sizes in $1D$. The curve of $\rho$ is for 
$L = 4000$ and those of $\rho_{surv}$ for from top to bottom $L = 20, 100$ and
$500$. The slope of the straight line is $\theta = 0.14$. In the inset, the 
same data in $2D$ are represented; $\rho_{surv}$ curves correspond from top to bottom
to $L = 10,25,70$ and $L = 280$ and $\rho$ to $L = 280$. The slope of the line
is $\theta = -0.65$ \protect\cite{compare}.}
\end{center}
\end{figure} 
We have studied systems of linear size up to $L = 4000$ in 
$1D$, $L = 400$ in $2D$ and $L = 80$ in $3D$. The dependence of the stationary
activity density on system size for several values of $\phi_0$ in a
one dimensional system is shown in Fig. 1. 
From this picture, we deduce the position of the critical point 
in $1D$, $\phi_c(1D) = 1.6371(1)$ (numbers in
parentheses correspond to the statistic uncertainty in the last digit). 
Also, from the slope of the log-log plot, we obtain $\beta/\nu_\perp(1D) = 0.213(6)$.
 The exponent $\beta$ may be estimated in an independent way from the scaling of the 
stationary value of $\rho_{surv}$ for large system sizes, as a function
of  $(\phi_0-\phi_c)$  above the critical point. 
This gives $\beta(1D) = 0.28(1)$. 
By studying the time evolution of the characteristic time of the
 surviving probability $P(t)$ at criticality, we obtain $z(1D) = 1.47(2)$. 
Finally, the exponent $\theta(1D) = 0.14(1)$ may be measured from the 
critical power law decay of $\rho$ in time, as may be seen in Fig. 2. 
Errors in these exponents mostly come from the uncertainty in the determination
 of the critical point. 
Repeating this process in higher dimensions, we find 
$\phi_c(2D) = 0.6325(5)$ and $\phi_c(3D) = 0.456(1)$, together with 
the critical exponents listed in Table I. 
In the table, we have also included the critical exponents of discrete
models claimed to belong to this same universality, and also 
(for comparison) those of the Directed Percolation (DP) class.
Observe the rather remarkable agreement (within errorbars) between all 
the measured exponents are their counterparts in discrete models.
Let us remark that, for those exponents for which the differences 
with DP-values are larger, our values also deviate from DP.
\begin{figure}
\begin{center}
\epsfxsize=6.5cm
\epsfbox{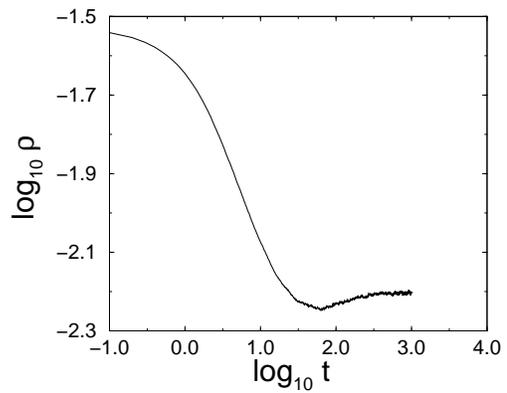}
\caption{Anomalous time decay of the activity density in $2D$, 
for $L = 280$ and $\phi_0 = 0.711$.}
\end{center}
\end{figure}
\begin{table}[b]
\begin{tabular}{clccccc}
   \hline
   \hline
   D$\ $ & Model$\ $ & $\beta$ & $\beta/\nu_\perp$ & $z$ & 
$\nu_\parallel$ & $\theta$ \\
   \hline
   $\,$ & Eq.(\ref{fesp})$\ \ $ &  $\ 0.28(1)\ $ & $\ 0.213(6)\ $ & 
   $\ 1.47(2)\ $ & $\ 1.63(1)\ $ & $\ 0.14(1)\ $ \\
   $1$ & C-DP &  $0.29(2)$ & $0.217(9)$ & $1.55(3)$ & $--$ & $0.140(5)$ \\
   $\,$ & DP &  $0.276...$ & $0.252...$ & $1.580...$ & $1.733...$ &
   $0.159...$ \\
   \hline
   $\,$ & Eq.(\ref{fesp})$\ \ $ &  $0.66(1)$ & $0.85(8)$ & $1.51(3)$ & 
   $1.27(7)$ & $0.50(5)$  \\
   $2$ & C-DP &  $0.64(2)$ & $0.78(2)$ & $1.55(3)$ & $1.29(8)$ & 
$0.51(1)$ \\
   $\,$ & DP &  $0.583(4)$ & $0.795(6)$ & $1.766(2)$ & $1.295(6)$ & 
$0.450(2)$ \\
   \hline
   $\,$ & Eq.(\ref{fesp})$\ \ $ &  $0.84(5)$ & $1.44(5)$ & $1.69(3)$ & 
$1.07(8)$ & $0.93(3)$ \\
   $3$ & C-DP &  $0.88(2)$ & $1.39(4)$ & $1.73(5)$ & $1.12(8)$ & 
$0.88(2)$ \\
   $\,$ & DP &  $0.80(2)$ & $1.39(1)$ & $1.901(5)$ & $1.105(5)$ & 
$0.730(4)$ \\
   \hline
   \hline
\end{tabular}
\label{tableI}
\caption{Critical exponents for ste\-ady state
   experiments in $d=1,2$ and $3$. Figures in parenthesis indicate the 
statistical uncertainty in the last digit. C-DP exponents are from
   Refs.~\protect\cite{Romu1,chate} and
   DP exponents from Refs.~\protect\cite{cp-dp}. In 1D, $\beta/\nu_\perp$ for C-DP
   has been calculated using the scaling relation $\beta/\nu_\perp = z \ \theta$.}
\end{table}
As we have already mentioned, some models in the Manna universality class may present
an anomalous behavior in the time decay of the activity density at the critical point 
\cite{fes,1d,Rossi,Romu1,Romu2}
$\rho(t,\phi_c)$.
This anomaly implies that, apparently, the scaling relation 
$\beta = \theta \, \nu_\parallel$  fails \cite{fes,1d,Rossi,Romu1,Romu2}, and 
that $\rho(t)$ may decay in a non-monotonous way at criticality.
In our case, there is no anomalous decay in $d=1$ (Fig.2). 
However, the anomaly is present both in $d=2$ and in $d=3$.
As can be seen in Fig. 3, the activity density decays initially faster than a power 
law, showing afterwards a non-monotonous behavior. 
The fast decay explains the anomalous values of $\theta$, not satisfying scaling 
relations, usually reported in the literature \cite{fes}.
Later on, $\rho$ increases before reaching the steady
state value, $\rho_{stat}(L)$ after a certain time $t_\times(L)$
(calculated by fixing some arbitrary criterion).
If the points $(t_\times(L),\rho_{stat}(L))$ 
are represented in a log-log plot at the
critical point, an alternative value for the exponent $\theta$ is found, 
which is related to the saturation time scale.
This value in two dimensional space, $\theta \approx 0.50$ (with a large 
statistic uncertainty), \cite{compare} is much closer 
to the more accurate measurements reported in the literature for 
models in this class ($\theta = 0.51$ for C-DP \cite{chate}). 
The common presence of anomalous behavior in discrete systems 
\cite{fes,1d,Rossi,Romu1,Romu2} and in the continuous theory
reinforces the claim that both belong to the same universality class: 
{\it they share not only the critical behavior but also the dynamical anomalies}. 
A deeper study of the physical origin of this anomaly is still missing but,essentially, 
it is related to the existence of different time scales.

Finally, let us mention that spreading experiments (performed using localized seeds
of activity \cite{MD,Hinrichsen}) exhibit some numerical instabilities, due
to the presence of large gradients, that merit a separate analysis and are,
therefore, left aside for a future work.

{\it Conclusions.--}  Strong numerical evidence confirms that there is a well defined
new universality class of systems with absorbing states, different from DP and
characterized by the presence of an extra static conserved field, linearly coupled
to the activity field. More importantly, the numerical simulations of the
phenomenological Langevin equation, proposed some time ago to capture the criticality
of this universality class, Eq.(\ref{fesp}), in one, two, and three dimensional systems, 
show that indeed it constitutes a sound minimal continuous representation of this class,
sharing all the critical exponents as well as the dynamical anomalies with the discrete
models. Therefore, no other higher order terms nor other noise correlations are needed 
to describe properly this class. Once the  situation has been clarified from the 
numerical side, further theoretical analyses are highly desirable in order  to put
this puzzling universality class under a more firm basis.  

\vspace{0.25cm}
We acknowledge M. A. Santos, H. Chat\'e, R. Pastor-Satorras, and R. Dickman for useful
comments, as well as P. Hurtado for his helpful participation 
in the early stages of this work. 
Support from the Spanish MCyT (FEDER) under project BFM2001-2841, 
and from the  postdoctoral program of
the ``Centro de F\'{\i}sica do Porto'' is acknowledged.
C. A. da Silva Santos acknowledges financial support from 
the Portuguese Research Council under grant SFRH/BPD/5557/2001.

\end{document}